\documentclass[prd,aps,superscriptaddress,eqsecnum,amsfonts,amsmath,twocolumn,preprintnumbers,floatfix,floats,nofootinbib,]{revtex4}
\pdfoutput=1
\usepackage[colorlinks=true, allcolors=red]{hyperref}
\usepackage{amsfonts,amsmath,units,wasysym,epsfig,graphicx,verbatim,color,
subfigure,epstopdf,bm}

\usepackage{siunitx}
\usepackage{multirow}

\usepackage[T1]{fontenc}

\definecolor{NTNUBlue}{rgb}{0.0470,0,0.5294}
\definecolor{Brown}{rgb}{0.5,0.5,0.5}
\definecolor{BrickRed}{cmyk}{0,.89,.94,.28}
\definecolor{FG}{cmyk}{0.75,0,1.0,0.5}

\DeclareSIUnit[number-unit-product = {}]
  \gauss{gauss}
\newcommand*{\gauss}{\, \mathrm{gauss}}
\newcommand*{\erg}{\, \mathrm{erg}}
\newcommand*{\ergs}{\erg}
\newcommand*{\second}{\, \mathrm{s}}
\newcommand*{\unsecond}{\mathrm{s}}
\newcommand*{\ms}{\, \mathrm{ms}}

\newcommand*{\hertz}{\, \mathrm{Hz}}

\newcommand*{\km}{\, \mathrm{km}}

\newcommand*{\kpc}{\, \mathrm{kpc}}
\DeclareSIUnit[number-unit-product = {}]
  \Mpc{Mpc}

\newcommand*{\euler}{\mathrm{e}}
\newcommand*{\ramuno}{\mathrm{i}}

\newcommand{\definiteIntegral}[4]{
	\int_{#1}^{#2} \! #3 \, \mathrm{d}#4
}

\newcommand{\derivative}[3][]{
	\ifthenelse{\isempty{#1}}{\ifthenelse{\isempty{#2}}{\frac{\mathrm{d}}{\mathrm{d} #3}}{\frac{\mathrm{d} #2}{\mathrm{d} #3}}}{\frac{\mathrm{d}^{#1} #2}{\mathrm{d} {#3}^{#1}}}
}

\newcommand{\partialLong}[3][]{
	\ifthenelse{\isempty{#1}}{\ifthenelse{\isempty{#2}}{\frac{\partial}{\partial #3}}{\frac{\partial #2}{\partial #3}}}{\frac{\partial^{#1} #2}{\partial {#3}^{#1}}}
}

\newcommand*{\unsim}{\mathord{\sim}}
\newcommand*{\angulardegree}{^{\circ}}

\newcommand*{\mathVector}[1]{\vec{\mathbf{#1}}}

\newcommand*{\tableref}[1]{TABLE~\ref{#1}}
\newcommand*{\figureref}[1]{FIG.~\ref{#1}}

\newcommand*{\sectionref}[1]{Sec.~\ref{#1}}

\newcommand*{\appendixref}[1]{Appendix \ref{#1}}

\newcommand*{\hrss}{h_\mathrm{rss}}

\newcommand*{\EGW}{E_\mathrm{GW}}

\newcommand*{\EEM}{E_\mathrm{EM}}

\newcommand*{\magneticfield}{\mathVector{B}}

\newcommand*{\Alfven}{Alfv\'{e}n}

\newcommand*{\bestUL}{4.3 \times 10^{46}}

\newcommand*{\bestULaLIGO}{9.0 \times 10^{44}}

\newcommand*{\bestULaLIGOclose}{3.2 \times 10^{43}}

\newcommand*{\bestULaLIGOcVsBurst}{3.2 \times 10^{-1}}

\newcommand*{\bestULaLIGOcVsGF}{1.9 \times 10^{-3}}

\newcommand*{\bestULSGR}{7.2 \times 10^{46}}

\newcommand*{\otherULSGR}{1.6 \times 10^{47}}

\newcommand*{\EGWEB}{0.43}

\newcommand*{\EGWEBSGR}{0.68}

\newcommand*{\EGWEBSGRother}{1.5}

\newcommand*{\EGWEMobserved}{5.9 \times 10^9}
\newcommand*{\EGWEMobservedSGRother}{2.1 \times 10^7}

\newcommand*{\T}{\rule{0pt}{2.6ex}}       
\newcommand*{\B}{\rule[-1.2ex]{0pt}{0pt}} 

\usepackage{amssymb}

\begin{document}
\newcommand{\rhat}{\hat{r}}
\newcommand{\iotahat}{\hat{\iota}}
\newcommand{\phihat}{\hat{\phi}}
\newcommand{\h}{\mathfrak{h}}
\newcommand{\be}{\begin{equation}}
\newcommand{\ee}{\end{equation}}
\newcommand{\ber}{\begin{eqnarray}}
\newcommand{\eer}{\end{eqnarray}}
\newcommand{\UO}{Department of Physics, University of Oregon, 1274 University of Oregon, Eugene, OR 97403-1274, U.S.A.}
\newcommand{\Harvard}{Department of Physics, Harvard University, 17 Oxford Street, Cambridge, MA 02138, U.S.A.}
\newcommand{\Monash}{School of Physics and Astronomy, Monash University, Clayton, Victoria 3800, Australia}
\newcommand{\Georgia}{Center for Relativistic Astrophysics, School of Physics, Georgia Institute of Technology, Atlanta, GA 30332, U.S.A.}
\newcommand{\Northwestern}{Department of Physics and Astronomy, Northwestern University, 2145 Sheridan Road, Evanston, IL  60208-3112, U.S.A.}
\newcommand{\OzGrav}{OzGrav: The ARC Centre of Excellence for Gravitational-wave Discovery, Hawthorn, Victoria 3122, Australia}

\newcommand\HL[1]{\textcolor{red}{#1}}

\title{Exploring a search for long-duration transient gravitational waves associated with magnetar bursts}

\author{Ryan Quitzow-James}
\affiliation{\UO}

\author{James Brau}
\affiliation{\UO}

\author{James Clark}
\affiliation{\Georgia}

\author{Michael~W.~Coughlin}
\affiliation{\Harvard}

\author{Scott~B.~Coughlin}
\affiliation{\Northwestern}

\author{Raymond Frey}
\affiliation{\UO}

\author{Paul Schale}
\affiliation{\UO}

\author{Dipongkar Talukder}
\affiliation{\UO}

\author{Eric Thrane}
\affiliation{\Monash}
\affiliation{\OzGrav}

\begin{abstract}
Soft gamma repeaters and anomalous X-ray pulsars are thought to be magnetars, neutron stars with strong magnetic fields of order $\unsim 10^{13}$--$10^{15} \gauss$. These objects emit intermittent bursts of hard X-rays and soft gamma rays.
Quasiperiodic oscillations in the X-ray tails of giant flares imply the existence of neutron star oscillation modes which could emit gravitational waves powered by the magnetar's magnetic energy reservoir.
We describe a method to search for transient gravitational-wave signals associated with magnetar bursts with durations of 10s to 1000s of seconds.
The sensitivity of this method is estimated by adding simulated waveforms to data from the sixth science run of Laser Interferometer Gravitational-wave Observatory (LIGO).
We find a search sensitivity in terms of the root sum square strain amplitude of $\hrss = 1.3 \times 10^{-21} \hertz^{-1/2}$ for a half sine-Gaussian waveform with a central frequency $f_0 = \SI{150}{\Hz}$ and a characteristic time $\tau = \SI{400}{\s}$. This corresponds to a gravitational wave energy of $\EGW = 4.3 \times 10^{46} \ergs$, the same order of magnitude as the 2004 giant flare which had an estimated electromagnetic energy of $\EEM = \unsim 1.7 \times 10^{46} (d/ \SI{8.7}{\kpc})^2 \ergs$, where $d$ is the distance to SGR 1806-20.
We present an extrapolation of these results to Advanced LIGO, estimating a sensitivity to a gravitational wave energy of $\EGW = \bestULaLIGOclose \ergs$ for a magnetar at a distance of $\SI{1.6}{kpc}$.
These results suggest this search method can probe significantly below the energy budgets for magnetar burst emission mechanisms such as crust cracking and hydrodynamic deformation.
\end{abstract}

\maketitle

\section{Introduction}\label{sec:introduction}

Laser Interferometer Gravitational-wave Observatory (LIGO) made the first direct detection of gravitational waves (GWs) from a merger of two black holes on September 14, 2015 \cite{discoveryPaper}.
GWs from a second black hole merger were detected on December 26, 2015 \cite{secondDetectionPaper}.
As more data is collected, searches will continue to be conducted for additional sources that might emit GWs.
One such potential source of GWs could be magnetars.

Soft gamma repeaters (SGRs) and anomalous X-ray pulsars (AXPs), astronomical objects that emit intermittent bursts of soft gamma rays and hard X-rays,
are thought to be magnetars, neutron stars with very strong magnetic fields estimated to be of order $\unsim 10^{13}$--$10^{15} \gauss$ \cite{DuncanThompson1992, MereghettiReview2015}.
The higher end of this range is $\unsim 1000$ times stronger than the estimated magnetic fields in normal pulsars typically of order $\unsim 10^{12} \gauss$ \cite{MereghettiReview, pulsarsLivingReview2008}.
A magnetar's magnetic field powers the electromagnetic bursts and quiescent emissions from these objects \cite{MereghettiReview, article:TheoreticalMagnetarReviewTurolla, MereghettiReview2015}.
The photon energies of hard X-rays and soft gamma rays in magnetar bursts range from above \SI{10}{keV} \cite{MereghettiReview} to hundreds of keV \cite{article:TheoreticalMagnetarReviewTurolla}.
Magnetar bursts could emit GWs at frequencies of interest (tens of hertz to kilohertz) to ground-based GW observatories by exciting nonradial modes such as f-modes, torsional modes and Alf\'{v}en modes \cite{S5MagnetarSearch, LevinVanHoven, shortHighFQPOs_1}.

There have been several previous searches for GWs associated with magnetar bursts.
One search targeted GWs emitted at the frequencies of modulations observed in the X-ray tail of the 2004 giant flare known as quasiperiodic oscillations (QPOs) \cite{hyperflareSearch}.
Three searches were sensitive to f-modes.
The first search was performed on data from the 2004 giant flare and on 190 burst events from the first year of LIGO's fifth science run (S5) \cite{firstFModeSearch}. The second search used a new method of stacking bursts \cite{flareStackingMethod} applied on data from the March 29, 2006 SGR 1900+14 storm \cite{magnetarBurstStormSearch}.
The third search was performed on data from the second year of S5, Virgo's first science run (VSR1), and the LIGO and GEO astrowatch period (A5) following S5 \cite{S5MagnetarSearch}.
In this paper, we describe a search pipeline for GW signals from magnetar bursts which extends the LIGO/Virgo searches to signals incorporating a wider range of signal morphologies \cite{RQJthesis}. We demonstrate the sensitivity of this search method using real data from LIGO's sixth science run (S6) and use this result to predict the sensitivity of this methodology for Advanced LIGO (aLIGO) \cite{aLIGOstandardReference}.

The remainder of this paper is organized as follows. In \sectionref{sec:magnetarburstsandgws}, we discuss different mechanisms for magnetar bursts and giant flares that can lead to the emission of gravitational radiation. In \sectionref{sec:methodology}, we develop the search method. We describe how the search sensitivity and the significance of an event are estimated.
Then, in \sectionref{sec:results}, we study the sensitivity of the search for various simulated GW signals.
We compare the estimated search sensitivity to different astrophysical parameters.
In \sectionref{sec:discussion}, we conclude by summarizing prospects for GW observations from magnetar bursts and giant flares.

\section{Magnetar Bursts and Gravitational Waves}\label{sec:magnetarburstsandgws}

Magnetar bursts are often grouped into three types:
short bursts lasting $\unsim0.1$--$\SI{1}{\s}$ with
isotropic energies up to $10^{41} \ergs$, intermediate bursts lasting $\unsim 1$--$\SI{40}{\s}$ with isotropic energies of $10^{41}$--$10^{43} \ergs$,
and giant flares lasting hundreds of seconds with isotropic energies of $10^{44}$--$10^{46} \ergs$ \cite{article:TheoreticalMagnetarReviewTurolla}.
In this paper we will show
that at aLIGO design sensitivity \cite{aLIGOdesignCurve}, a GW from a magnetar burst at a distance of order $\unsim 1 \kpc$ could be detectable with an energy of order $\unsim 10^{43} \ergs$.

Short and intermediate bursts have been observed from both SGRs and AXPs \cite{article:TheoreticalMagnetarReviewTurolla}.
Three giant flares have been observed from SGRs: the first on March 5, 1979, from SGR 0526-66, the second on August 27, 1998, from SGR 1900+14, and the third on December 27, 2004, from SGR 1806-20 \cite{MereghettiReview}.
For reviews of magnetars, see referenced articles on the observational properties of magnetars \cite{MereghettiReview}, the physics behind observations of magnetars \cite{article:TheoreticalMagnetarReviewTurolla}, and a more recent review of observational properties of magnetars and possible formation models \cite{MereghettiReview2015}.

\begin{figure}[h]

\centering
\includegraphics[width=\linewidth]{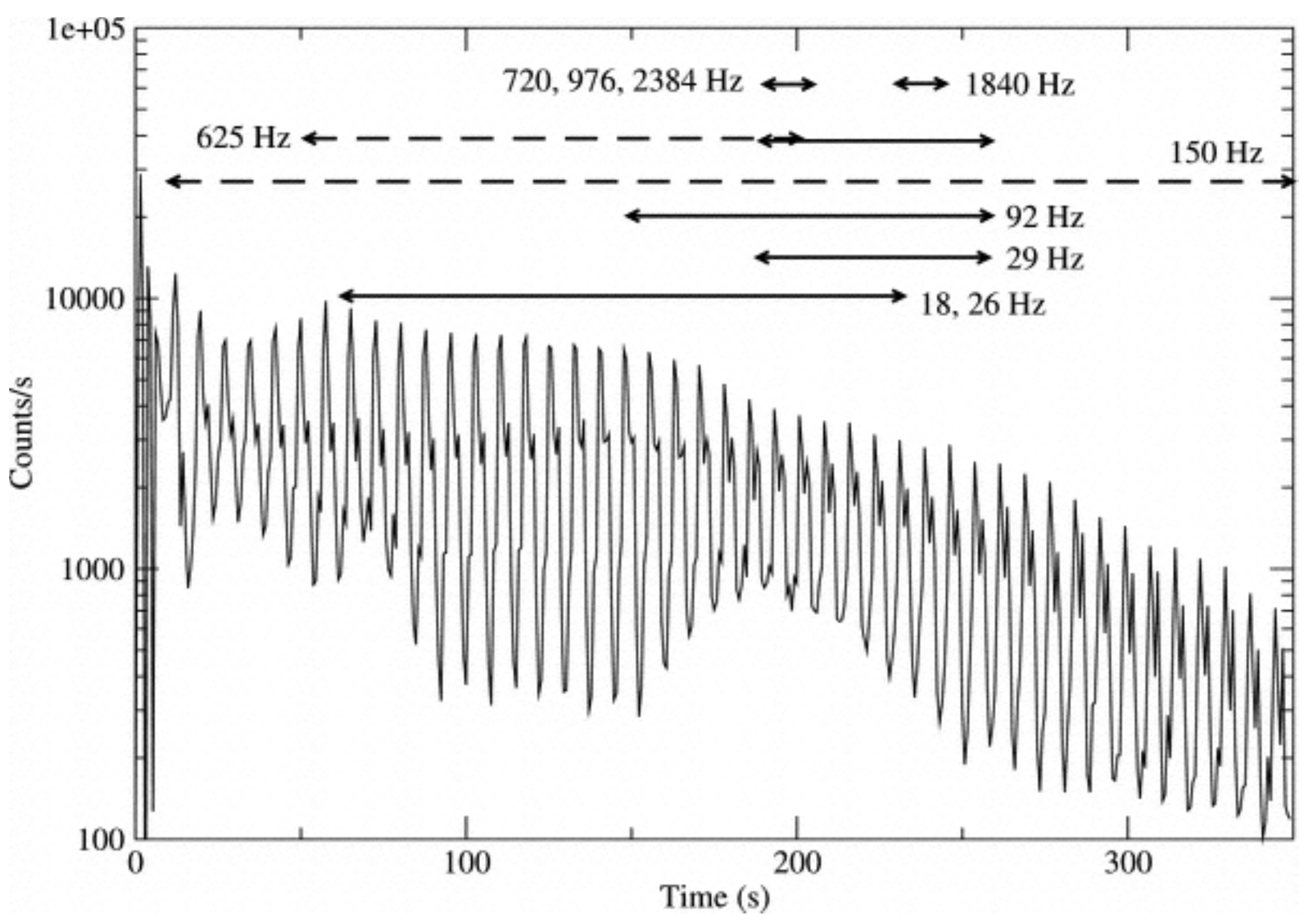}
\caption[Time series of data of the 2004 giant flare from SGR 1806-20.]{Time series of data of the 2004 giant flare from SGR 1806-20. Initial peak followed by X-ray tail. Large modulations in the tail are due to the magnetar rotation. Solid and dashed lines indicate detection of QPOs at different phases of the rotation. Figure \copyright{} AAS. Reproduced with permission from Ref~\cite{HyperflareGlobalTorsionalVibrations}.}
\label{figure:Hyperflare}
\end{figure}

Giant flares are characterized by an initial spike with luminosities up to $10^{47} \erg/\unsecond$ lasting less than a second, followed by an afterglow lasting several hundred seconds modulated by the rotation period of the neutron star \cite{article:TheoreticalMagnetarReviewTurolla}.
Higher frequency modulations known as quasiperiodic oscillations (QPOs) have been observed in the X-ray tails of giant flares (see \figureref{figure:Hyperflare}) \cite{MereghettiReview}; they were first discovered in the tail of the giant flare from SGR 1806-20 \cite{QPOdiscovery}.
Two more recent analyses revealed the possible existence of QPOs in some short burst events, showing that QPOs may not be limited to giant flares \cite{shortHighFQPOs_1, shortHighFQPOs_2}.

Potential causes of QPOs include torsional modes
and \Alfven{} modes, both of which have low frequencies that are hypothesized to be related to QPOs due to the overlap in frequencies \cite{Duncan1998, QPOdiscovery, HyperflareGlobalTorsionalVibrations, shortHighFQPOs_1, LevinVanHoven, ThreeEvolutionaryPaths}.
If either of these modes is responsible for QPOs, they might emit GWs for longer durations than the visible QPOs in the electromagnetic spectrum \cite{ThreeEvolutionaryPaths}. Some QPOs have been observed lasting hundreds of seconds (see \figureref{figure:Hyperflare}), which could translate to GWs with durations of hundreds to thousands of seconds \cite{ThreeEvolutionaryPaths, HyperflareGlobalTorsionalVibrations}.
If the GW lasts as long as the observed QPO, then the duration
could be tens to hundreds of seconds. K. Glampedakis and D. I. Jones \cite{ThreeEvolutionaryPaths} speculate a wide range of duration for \Alfven{} modes that can accommodate observed durations of QPOs.
They also suggest that \Alfven{} mode frequencies may evolve over time by increasing in frequency \cite{ThreeEvolutionaryPaths}.

Multiple mechanisms have been put forth as possible causes of magnetar bursts and giant flares.
These include magnetic reconnection, crust cracking and hydrodynamic deformation.
Magnetic reconnection can occur when the magnetic field reaches an unstable configuration \cite{ThompsonDuncan1995}.
In the crust cracking scenario, the decay of the interior magnetic field builds up stress in the crust until it breaks. The crust then moves by rapid plastic deformation and twists the external magnetic field lines, injecting energy into the magnetosphere \cite{article:TheoreticalMagnetarReviewTurolla}. This can also cause magnetic reconnections depending on the configurations of the evolving magnetic field lines.
In the case of hydrodynamic deformation \cite{article:Ioka, CorsiOwenEnergyEstimates}, the magnetic field in the core has a strong toroidal component which makes the star's shape more prolate.
Jumps between equilibrium configurations of the magnetic field release gravitational potential energy to power the bursts by reducing the toroidal component, making the the star less prolate \cite{CorsiOwenEnergyEstimates, article:Ioka}.

Although the exact mechanisms behind bursts are uncertain, as are their couplings to GWs, energy considerations can be used to estimate the strength of possible GWs.
Simulations by Horowitz and Kadau \cite{HorowitzKadau} suggest that the breaking strain of the neutron star crust is much stronger than previously thought, possibly generating energies up to $\unsim 10^{46} \erg$, enough to power the observed giant flares  \cite{CorsiOwenEnergyEstimates}.
Energies up to $10^{49}$--$10^{50} \erg$ may be generated if the neutron star (or just the core) is made up of solid quark matter \cite{CorsiOwenEnergyEstimates}.
Hydrodynamic deformation could generate energies up to $\unsim 10^{49} \ergs$ \cite{CorsiOwenEnergyEstimates}.
For sources in the distance range of known magnetars of 1.6--\SI{62.4}{kpc} \cite{mcgillMagnetarList}, both crust cracking and hydrodynamic deformation could provide sufficient energy to generate GWs with amplitudes comparable to the current sensitivities of ground-based gravitational wave observatories.

Since the exact astrophysical mechanisms behind magnetar bursts are unknown, a search method sensitive to a wide variety of possible signals is needed.
Furthermore, observations of magnetar burst afterglows, specifically QPOs, motivate a search method which is sensitive to long-duration, narrowband transient signals, including nearly monochromatic signals.
Levin and van Hoven \cite{LevinVanHoven} compute an estimate which finds that f-modes may be less excited than torsional modes, giving additional motivation to search for GWs at QPO frequencies.
Previous searches for GWs from magnetars focused on short duration signals of the order \SI{100}{\ms} emitted from f-modes \cite{S5MagnetarSearch}.
A search for long-duration transient GWs associated with the 2004 giant flare was performed using the observed frequencies and durations of QPOs in the giant flare's afterglow \cite{hyperflareSearch}.
The search calculated excess power in bands of order $\unsim \SI{10}{\Hz}$ around the QPO frequencies for durations up to \SI{350}{\s} \cite{hyperflareSearch}.
This search method differs from the short duration searches by looking for long-duration transient GW signals, and differs from the 2004 giant flare GW search because it does not require a specific search frequency, makes no assumptions regarding the frequency content, and is sensitive to other signal models (long-duration transient monotonic narrowband signals that can increase or decrease in frequency \cite{seedlessPaper}).

\section{Method}\label{sec:methodology}\label{sec:methodology}

Electromagnetic observations provide times and sky locations of magnetar burst events. Since the existing models do not predict the details of GWs, we search for unmodelled GW signals by looking for correlated patterns in the data of two (or more) detectors after accounting for time delays and detector responses consistent with a given sky location. Data is transformed into spectrograms or time-frequency maps (tf-maps) to look for excess power in time-frequency pixels (tf-pixels).
A targeted search such as this is more sensitive than an all-time, all-sky search because the inclusion of time and sky position reduces the probability that random noise fluctuations and instrumental artifacts will conspire to produce a false detection.
While the waveforms of the GWs are unknown, we assume GWs with behavior similar to that of the QPOs when assessing the search sensitivity.

Several pipelines could be used to search for GW signals associated with QPOs. The excess power method used in Ref \cite{hyperflareSearch} and later extended using a multi-trigger concept and the ability to combine data from multiple detectors and tested on Gaussian data and S5 playground data \cite{hyperflareMethodPlaygroundDataAnalysis} is one possibility.
Other methods currently used to search for different GW signals which could be adapted for this source include machine learning algorithms used in a sensitivity study on r-modes in newborn neutron stars \cite{rmodeNSSensitivityStudy} and a flexible cross-correlation method \cite{crossCorrelationMethod} which has been tuned for intermediate duration GW signals \cite{crossCorrelationMethodIntermediate}.

We utilize the Stochastic Transient Analysis Multi-detector Pipeline (STAMP) \cite{STAMPmethodsPaper} to develop a search for GWs from magnetar bursts.
STAMP was previously used to search for long-lived GW transients coincident with long gamma-ray bursts \cite{longGRBsearch} and in an all-sky search for long-duration GW transients \cite{STAMPallSky}.

\subsection{Search Pipeline: STAMP}\label{stamp_section}

STAMP calculates an estimator $\hat{Y}$ of the cross-power of two detectors for sky direction $\hat{\Omega}$ in a single tf-pixel starting at time $t$ for duration $\delta t$ at frequency $f$ with frequency resolution $\delta f$ given by \cite{STAMPmethodsPaper}
\begin{equation}
\hat{Y}(t;f,\hat{\Omega}) \equiv \mathrm{Re} \left[2 \tilde{Q}_{IJ}(t;f,\hat{\Omega}) \tilde{s}^*_I(t;f) \tilde{s}_J(t;f) \right]\;,
\end{equation}
where $\tilde{Q}_{IJ}(t;f,\hat{\Omega})$ is a sky-position-dependent time-frequency filter function that accounts for the source sky location, polarization and the detector antenna responses.
$\tilde{s}_I(t;f)$ denotes the short-term Fourier transform over $\delta t$ of the power series $s_I(t; f)$ for detector $I$ for the pixel starting at time $t$ and frequency $f$.
For an unpolarized source, the filter function can be expressed as \cite{STAMPmethodsPaper}
\begin{equation}
\tilde{Q}_{IJ}(t;f,\hat{\Omega}) = \frac{\euler^{2\pi \ramuno f \hat{\Omega} \cdot \Delta \overrightarrow{x}_{IJ}/c }}{\frac{1}{2} \sum_A F^A_I(t;\hat{\Omega}) F^A_J(t;\hat{\Omega})}\;,
\end{equation}
where $I$ and $J$ refer to the detectors, $F^A_I(t;\hat{\Omega})$ is the antenna response \cite{antennaResponseAllenRomano} for detector $I$ and $\Delta \overrightarrow{x}_{IJ} \equiv \overrightarrow{x}_I - \overrightarrow{x}_J$ is the difference in position vectors of the detectors. $A$ is summed over plus ($+$) and cross ($\times$) GW polarizations.

An estimator for the variance of $\hat{Y}$ is \cite{STAMPmethodsPaper} given by
\begin{equation}
\hat{\sigma}^2_Y(t;f,\hat{\Omega}) = \frac{1}{2} | \tilde{Q}_{IJ}(t;f,\hat{\Omega}) |^2 P^{\mathrm{adj}}_I(t;f) P^{\mathrm{adj}}_J(t;f)\;,
\end{equation}
where $P^{\mathrm{adj}}_I(t;f)$ is the one-sided auto-power spectrum of the $I$th detector averaged over neighboring pixels.

A tf-map is created in which each pixel has a value calculated from this statistic. The signal to noise ratio (SNR) from STAMP for a single pixel is then defined as \cite{STAMPmethodsPaper}
\begin{equation}
\mathrm{SNR}(t;f,\hat{\Omega}) \equiv \frac{\hat{Y} (t;f,\hat{\Omega})}{\hat{\sigma}_Y(t;f,\hat{\Omega})}\;.
\end{equation}
After the filter function is applied to the tf-map, pixels are grouped into clusters and analyzed to search for possible signals.
Clusters are groups of pixels for which a combined SNR is calculated when searching for a potential signal.
For this method, we use \textit{stochtrack}, a seedless clustering algorithm looking for long-duration transient narrowband monotonic signals \cite{seedlessPaper}.
Many clustering algorithms generate clusters using seeds, pixels with an SNR above a given threshold.
A long-duration transient GW produces less excess power in an individual pixel than a short burst of the same total energy, making it less likely to produce seed pixels to form a cluster for a seed-based clustering algorithm to make a statistically significant detection \cite{seedlessPaper}.
Stochtrack was created with this in mind and was found to significantly improve sensitivity to these signals in comparison to two seeded clustering algorithms \cite{seedlessPaper}.

Stochtrack uses quadratic B\'{e}zier curves to trace tracks through the tf-map which group pixels into clusters \cite{seedlessPaper}.
Stochtrack then finds the loudest cluster.
The randomly generated curves persist for a minimum time $t_{min}$ and are generated with three time-frequency points: $P_0 (t_{\mathrm{start}}, f_{\mathrm{start}})$, $P_1 (t_{\mathrm{mid}}, f_{\mathrm{mid}})$ and $P_2 (t_{\mathrm{end}}, f_{\mathrm{end}})$. These points are used to form a quadratic B\'{e}zier curve parameterized by $\xi = [0,1]$ \cite{seedlessPaper}:
\begin{equation}
\left( \begin{array}{c} t(\xi) \\ f(\xi) \end{array} \right) = (1-\xi)^2 P_0 + 2(1-\xi)\xi P_1 + \xi^2 P_2\;.
\end{equation}
It should be noted the quadratic B\'{e}zier curve is an approximate fit to an arbitrary monotonic curve, and that it may be a poor fit for broadband or non-monotonic signals \cite{seedlessPaper}.

In a previous study \cite{seedlessPaper}, $2 \times 10^7$ clusters were found to provide remarkable sensitivity. The computational time was found to increase linearly with the number of trials \cite{seedlessPaper}. We increased the number of clusters used to $3 \times 10^7$, which maintained a decent SNR sensitivity on the map size we used, while still having a reasonable computation time.

The single-pixel SNR can be generalized to calculate the SNR of a cluster of $N$ pixels which form a set of pixels, $\Gamma$ \cite{STAMPmethodsPaper}:
\begin{equation}\label{clusterSNR}
\mathrm{SNR}_\Gamma(\hat{\Omega}) = \left| \frac{\sum_{t;f \in \Gamma} \mathrm{SNR}(t;f,\hat{\Omega})}{\sqrt{N}} \right| \;.
\end{equation}
\eqref{clusterSNR} is based on the multi-pixel statistic derived in Ref~\cite{STAMPmethodsPaper}, but uses a different normalization based on the number of pixels in the cluster.
When the B\'{e}zier curve defining the cluster goes through multiple tf-pixels occurring during the same time segment $t + \delta t$, the SNR of each pixel is weighted by the fraction of the pixel time duration ($\delta t$) that the curve is in the pixel.

We take the absolute value in \eqref{clusterSNR} because at certain sky positions, some polarizations interacting with the unpolarized filter function produce negative SNR.
At some sky positions, the plus or cross GW polarizations are anti-correlated in the two detectors due to the slightly different orientations of the detectors. 
When only one polarization is anti-correlated, the cross-power switches from positive to negative as the polarization changes, while sign of the filter function stays the same.
This means some polarizations will produce negative SNR.

\subsection{Data}\label{data}

We search for signals in GW data in an \textit{on-source} window of $[-2, 1600]$~s around the reported magnetar burst time.\footnote{STAMP's implementation of 50\% overlapping pixels adds \SI{2}{\s} to the end of the on-source window, making the effective on-source window $[-2, 1602]$~s.} The \SI{2}{\s} stretch preceding the trigger accounts for timing uncertainties from the satellites used for the electromagnetic observations, as well as the difference in time recorded at the satellites and on Earth. The \SI{1600}{\s} following the trigger is based on two factors: 1) observations that the longest afterglow from a giant flare was approximately \SI{400}{\s} \cite{MereghettiReview}, and 2) the possibility that GW signals associated with QPOs may last several times longer than the observed QPOs in the electromagnetic afterglow of the giant flares \cite{ThreeEvolutionaryPaths, HyperflareGlobalTorsionalVibrations}.
To avoid artifacts, the data is processed initially using a wider window of [-20, 1620] s.
We use $\SI{4}{\s} \times \SI{1}{\Hz}$ tf-pixels to generate the tf-map for this window as this is the minimum resolution for which we can calculate such a large window due to computational limits.

We use data from LIGO's two detectors, LIGO Hanford Observatory (LHO) and LIGO Livingston Observatory (LLO), during S6 to estimate the sensitivity of this search method \cite{S6dataRelease}.
The search bandwidth considered here is 40--\SI{2500}{\Hz}.
This is driven by the sensitive frequency band of LIGO during S6.
The lower frequency bound comes from seismic noise while shot noise gradually limits sensitivity at higher frequencies (see \figureref{fig:hrssestimate}).
This range accommodates the observed frequencies of QPOs.
The higher bound is chosen to include the highest observed QPO frequencies in the electromagnetic afterglows of the giant flares, including the QPO at \SI{2384}{\Hz} \cite{MereghettiReview}.

This study used data surrounding magnetar burst triggers that occurred when both LIGO detectors were operational with science quality data.
The trigger times and source objects were obtained from the InterPlanetary Network (IPN), using the IPN master burst list \cite{sgrList}. Three magnetar bursts occurred while the LIGO detectors were active and recording data during S6.
The first two bursts, numbered 2469 and 2471 on the burst list, were from SGR 1806-20 (an SGR). The third burst, numbered 2475, was from 1E 1841-045 (an AXP).
The sky positions and distances of these objects were provided from the McGill Online Magnetar Catalog \cite{mcgillMagnetarList, McGillMagnetarListPaper}.\footnote{The McGill Online Magnetar Catalog is available available online at \url{http://www.physics.mcgill.ca/~pulsar/magnetar/main.html} \cite{mcgillMagnetarList} and in Ref.~\cite{McGillMagnetarListPaper}.}
The estimated distances to SGR 1806-20 and 1E 1841-045 are \SI{8.7}{kpc} and \SI{8.5}{kpc} respectively~\cite{mcgillMagnetarList}.

The detectors sometimes have decreased sensitivity due to environmental noise or other factors.
We use identical data quality cuts to those used in the all-sky long-transient search \cite{STAMPallSky}, removing time segments during which identified instrumental or environmental noise sources coupled to the GW strain signal as well as the times when hardware injections were present. We remove 2.2\% of LHO and LLO coincident data from S6 as potential data to analyze \cite{STAMPallSky}.
STAMP also utilizes an auto-power consistency cut between the detectors \cite{Xi_cut_paper}.

In addition, the sensitivity of a cross-correlation search could be affected by elevated noise in particular frequencies. Frequencies that appear with an identified source in the online notch list \cite{notchlist} were notched and removed from the analysis. Additionally, a frequency near a \SI{16}{\Hz} harmonic, frequencies adjacent to \SI{60}{\Hz} harmonics of power lines, \SI{372}{\Hz} (identified as a \SI{2}{\Hz} harmonic), and a few frequencies adjacent to pulsar injections \cite{pulsarInjectionPaper} also had to be notched.
The \SI{2}{\Hz} and \SI{16}{\Hz} harmonics were believed to be due to the use of high-frequency ($> \SI{1}{\kHz}$) dither signals causing LIGO's digital-to-analog converters (DACs) to generate low frequency noise \cite{S6DetcharPaper}.

\subsection{Background Estimation}

In a search for GW signals triggered by an electromagnetic counterpart, the significance of an event in the on-source is determined by estimating the background noise using \textit{off-source} data,
data from times before and after the on-source window.
To prevent mistaking a possible signal from an unknown source as noise, the background is estimated from data shifted to non-physical time delays, times longer than the light crossing time of the detectors.

One way to time shift to non-physical times is to pair off-source windows from different times.
This ensures that any time delay is non-physical.
This also allows more detector pair data sets to be generated from less data, as a single time segment from one detector can be paired with multiple time segments from the other detector.
Using less data allows the time segments to be focused closer to the on-source window.

We use 33 non-overlapping time segments from each detector when both detectors are active and recording data to generate 1000 time shifted off-source data segment pairs to estimate the background for each magnetar burst. With 1000 segment pairs, the probability of noise creating an SNR louder than the background in any of the 3 magnetar bursts' on-source windows is $~0.3\%$, which corresponds to a significance of nearly 3-sigma. The off-source time-frequency window of each time segment has the same time duration and frequency bandwidth as the on-source window. The segments are \SI{1640}{s} initially to avoid artifacts and processed into \SI{1602}{s} (\SI{1604}{s} due to the overlapping pixels) tf-maps like the on-source window. The segments are taken from data as close to the on-source window as possible.

The loudest cluster SNRs of the off-source time shifted data segment pairs are combined into a background SNR distribution.
This is used to estimate the false alarm probability (FAP) of the loudest cluster in the on-source window of a trigger.
The FAP of a cluster
is estimated by the ratio of clusters in the background with SNR greater than or equal to that cluster's SNR ($N_{\geqslant}$) to the total number of clusters in the background ($N_{\mathrm{Total}}$):
\begin{equation}
\mathrm{FAP} = N_{\geqslant}/N_{\mathrm{Total}}\;.
\end{equation}

\begin{figure}[h!]
{\includegraphics[width=\linewidth]{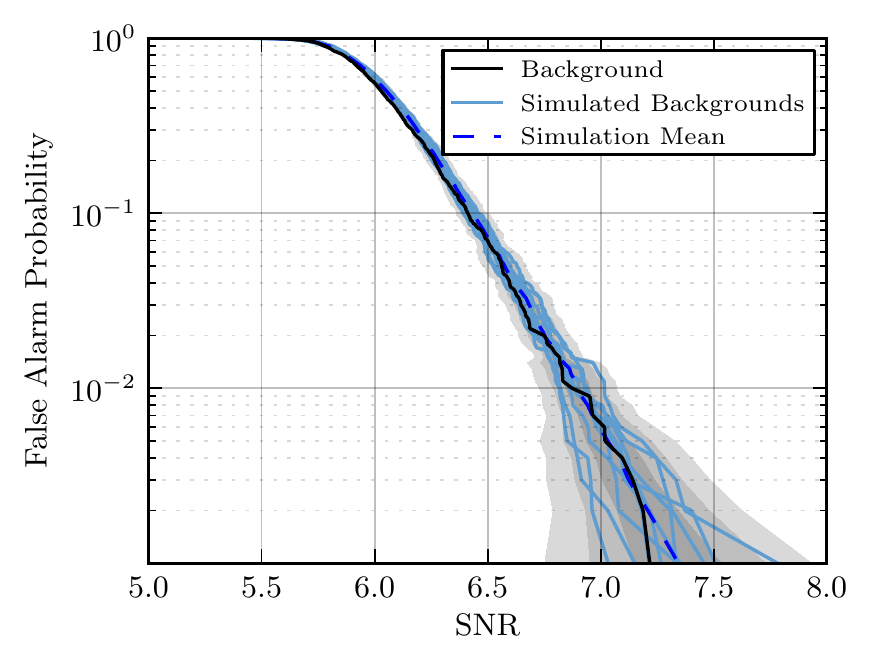}}
  \caption{The background SNR distribution for magnetar burst 2471. The shaded grey areas mark the regions within $\sigma = 1, 2, 3$ thresholds along the x-axis. The variance $\sigma$ was calculated for each false alarm probability using the SNRs from 10 simulated SNR distributions.}
\label{fig:background}
\end{figure}

To gauge the data quality of the noise background, the background SNR distribution can be compared to simulated background SNR distributions (see \figureref{fig:background}).
See Ref.~\cite{fiveSigmaPaper} for a thorough characterization of using simulated data to estimate the background distribution.
For this study, we compare each background SNR distribution to 10 simulated SNR background distributions. Each simulated background SNR distribution is generated using 1000 simulated segment pairs. Thirty-three sets of simulated segments are generated for each detector and used to construct the 1000 segment pairs. Simulated segments are made using Gaussian data weighted by the on-source power spectral density for the associated detector.

\subsection{Detection Efficiency}\label{upperLimitSection}

The mechanisms behind the magnetar bursts and their coupling to GWs are uncertain.
Models for magnetar bursts and giant flares do not yet predict specific GW waveforms related to QPOs.
We therefore use electromagnetic QPO observations as a guide to create ad-hoc GW waveforms for magnetar bursts.
As actual QPOs can vary in both frequency and amplitude, we simulate plausible long-duration, damped oscillation signals following a sudden burst using a smoothly tapered half sine-Gaussian ($t~\geqslant~0$ in \eqref{fullHplusSG} and \eqref{fullHcrossSG}) as a rough approximation.
This half sine-Gaussian is used instead of a full sine-Gaussian because it has
a large initial amplitude which damps over time.

The GW strain for the plus ($h_+$) and cross ($h_\times$) polarizations of a sine-Gaussian with amplitude $h_0$, central frequency $f_0$, characteristic time $\tau$, polarization angle $\psi$, and inclination angle $\iota$ are given by
\begin{multline}
h_+ = \frac{h_0}{\sqrt2}\euler^{-t^2/\tau^2}\left(\frac{\left(1 + \cos^2\iota\right)}{2} \cos{\left(2\,\pi f_0 t\right)} \cos{2\psi}\right. \\ \left.\vphantom{\frac{\left(1 + \cos^2\iota\right)}{2}} + \left(\cos{\iota}\right) \sin{\left(2\,\pi f_0 t\right)}\sin{2\psi}\right)\;,\label{fullHplusSG}
\end{multline}
\begin{multline}
h_\times = \frac{h_0}{\sqrt2}\euler^{-t^2/\tau^2}\left(-\frac{\left(1 + \cos^2\iota\right)}{2} \cos{\left(2\,\pi f_0 t\right)} \sin{2\psi}\right. \\ \left. \vphantom{\frac{\left(1 + \cos^2\iota\right)}{2}} + \left(\cos{\iota}\right) \sin{\left(2\,\pi f_0 t\right)}\cos{2\psi}\right)\;.\label{fullHcrossSG}
\end{multline}
The inclination factors of $\left(1 + \cos^2\iota\right)/2$ and $\left(\cos{\iota}\right)$ come from the mass quadrupole, the lowest order emission mode of GWs. The inclination angle $\iota$ is the angle between the source sky location and the axis of rotation of the quadrupole.

To estimate the search sensitivity, we add simulated GW signals into data and apply the analysis method.\footnote{In order to facilitate rapid calculation of the detection efficiency, we employ a technique in which we find the stochtrack template (or small set of templates) that best matches the injected signal and use that to quickly calculate the SNR \cite{RQJthesis}.}
Half sine-Gaussians are injected into 40 off-source segment pairs to estimate a 90\% detection efficiency at $\mathrm{FAP}=10^{-3}$, found by recovering 90\% of the waveforms.
A waveform is recovered if its SNR is greater than a threshold SNR, in this case the loudest SNR from the background distribution.\footnote{When finding injected waveforms, the pixels making up the loudest cluster in each segment pair from the background are set to zero to avoid contamination from the loudest cluster.}
The amplitude ($h_0$) is varied until at least 90\% of the waveforms are recovered.

High frequency QPOs are thought to damp in less than \SI{1}{\s} \cite{shortHighFQPOs_1, shortHighFQPOs_2}, so this sensitivity study focused on low frequencies.
The frequency of \SI{150}{\Hz} was chosen to cover the most sensitive band of the detectors, and the frequencies of \SI{450}{\Hz} and \SI{750}{\Hz} were chosen to cover the lower-mid frequency range outside of the most sensitive frequency band.
Two values of the characteristic time $\tau$ were chosen to cover both shorter and longer signals: $\tau = \SI{150}{\s}$ and $\tau = \SI{400}{\s}$.
The length of injected waveforms are $3\tau$.
The first and second halves of a \SI{2}{\s} Hann window were applied to beginning and end of the simulated signal respectively.

The inclination angle of the GW source will affect the strain at the detector (see \eqref{fullHplusSG} and \eqref{fullHcrossSG}). We focus on the polarization this search method is most sensitive to: circularly polarized waveforms (with inclination angle $\iota = 0\angulardegree$ and polarization angle $\psi = 0\angulardegree$).

\section{Search Sensitivity}\label{sec:results}

We estimated the search sensitivity using simulated waveforms injected into the background data of magnetar burst events, increasing the amplitude until the waveforms were found with a 90\% detection efficiency.
The results for circularly polarized signals (with $\iota = 0\angulardegree$ and $\psi = 0\angulardegree$) are summarized in \tableref{table:upperLimitsCircular}.
The root sum square strain ($\hrss$) and the GW energy ($\EGW$) of simulated signals were calculated as detailed in \appendixref{upper_limits}.
The estimated source distances from the McGill Online Magnetar Catalog \cite{mcgillMagnetarList} were used to calculate the GW energy.
These results include the calibration error (see \appendixref{cal_err} for details), which leads to a 15\%--21\% increase in the GW strain of the recovered waveforms (32\%--46\% increase in energy).

\begin{table}[h]
\centering

\begin{tabular}{*{5}{c}}
	\hline
	\T \B $f_0$ & $\tau$ & $h_{rss}$ & Distance & $E_{\mathrm{GW}}$ \\
	\hline
	\T \B \si{\Hz} & \si{\s} & $\si{\Hz}^{-1/2}$ & kpc & $\ergs$ \\

	\hline
    	\multicolumn{5}{| c |}{SGR trigger 2469} \T \B \\
    	\hline

	\T $150$ & \multirow{3}{*}{$400$} & $1.7 \times 10^{-21}$ & \multirow{6}{*}{$8.7$} & $7.2 \times 10^{46}$ \\

	$450$ &  & $2.8 \times 10^{-21}$ &  & $1.8 \times 10^{48}$ \\

	\B $750$ &  & $3.5 \times 10^{-21}$ &  & $8.0 \times 10^{48}$ \\

	\cline{1-3}\cline{5-5}

	\T $150$ & \multirow{3}{*}{$150$} & $3.0 \times 10^{-21}$ &  & $2.3 \times 10^{47}$ \\

	$450$ &  & $4.3 \times 10^{-21}$ &  & $4.2 \times 10^{48}$ \\

	\B $750$ &  & $5.2 \times 10^{-21}$ &  & $1.8 \times 10^{49}$ \\
	\hline
    	\multicolumn{5}{| c |}{SGR trigger 2471} \T \B \\
    	\hline

	\T $150$ & \multirow{3}{*}{$400$} & $2.5 \times 10^{-21}$ & \multirow{6}{*}{$8.7$} & $1.6 \times 10^{47}$ \\

	$450$ &  & $5.3 \times 10^{-21}$ &  & $6.5 \times 10^{48}$ \\

	\B $750$ &  & $6.4 \times 10^{-21}$ &  & $2.6 \times 10^{49}$ \\

	\cline{1-3}\cline{5-5}

	\T $150$ & \multirow{3}{*}{$150$} & $8.0 \times 10^{-21}$ &  & $1.7 \times 10^{48}$ \\

	$450$ &  & $1.3 \times 10^{-20}$ &  & $3.9 \times 10^{49}$ \\

	\B $750$ &  & $1.6 \times 10^{-20}$ &  & $1.6 \times 10^{50}$ \\
	\hline
    	\multicolumn{5}{| c |}{SGR trigger 2475} \T \B \\
    	\hline

	\T $150$ & \multirow{3}{*}{$400$} & $1.3 \times 10^{-21}$ & \multirow{6}{*}{$8.5$} & $4.3 \times 10^{46}$ \\

	$450$ &  & $2.2 \times 10^{-21}$ &  & $1.1 \times 10^{48}$ \\

	\B $750$ &  & $3.3 \times 10^{-21}$ &  & $6.8 \times 10^{48}$ \\

	\cline{1-3}\cline{5-5}

	\T $150$ & \multirow{3}{*}{$150$} & $4.5 \times 10^{-21}$ &  & $5.0 \times 10^{47}$ \\

	$450$ &  & $5.0 \times 10^{-21}$ &  & $5.5 \times 10^{48}$ \\

	\B $750$ &  & $6.6 \times 10^{-21}$ &  & $2.7 \times 10^{49}$ \\
	\hline
\end{tabular}

\caption{Search sensitivity for 90\% detection efficiency for circularly polarized signals with $\iota$ $=$ $0\angulardegree$ and $\psi$ $=$ $0\angulardegree$.}
\label{table:upperLimitsCircular}
\end{table}

The best estimated search sensitivity was to a circularly polarized half sine-Gaussian with central frequency $f_0 = 150 \hertz$ and characteristic time $\tau = 400 \second$. This waveform was recovered with an energy $\EGW = \bestUL \erg$ for magnetar burst 2475 from 1E 1841-045. For bursts from SGR 1806-20, it was recovered for magnetar burst 2469 with an energy $\EGW =  \bestULSGR \erg$ and for magnetar burst 2471 with an energy of $\EGW = \otherULSGR \erg$. As the duration of $\tau$ increases, this method is sensitive to waveforms with lower $\hrss$ and $\EGW$.

$\EGW$ can be compared to the estimated magnetar burst energies ($\EEM$) from electromagnetic observations.\footnote{The magnetar burst library at \url{http://staff.fnwi.uva.nl/a.l.watts/magnetar/mb.html} was used to find the paper containing the estimated electromagnetic energy for magnetar burst 2475 \cite{magnetarBurstLibrary}.}
Electromagnetic energy estimates were available for magnetar burst 2471 and 2475, but were unavailable for magnetar burst 2469 and we have marked it as unknown in \tableref{table:upperLimitsAndResultsSummary}. Magnetar burst 2471 had an estimated fluence of $7.88 (\pm 0.39) \times 10^{-7} \ergs/\si{\cm}^2$ in the 8--1000 keV band and lasted $\unsim \SI{100}{\ms}$ \cite{GCNcircularSGR180620burst, FiveYearFermiCatalog}. Assuming isotropic emission and using the estimated distance to SGR 1806-20 of 8.7 kpc, this fluence estimate corresponds to a burst energy of $7.14\times10^{39}\ergs$.
Magnetar burst 2475 was the first detected burst from 1E 1841-045, with a duration of \SI{32}{\ms} and an estimated energy in the 15--100 keV band of $7.2^{+0.4}_{-0.6} \times 10^{36} \ergs$ \cite{AXPburstEMenergy}.
Using the best search sensitivity for each burst and ignoring the uncertainties on the $\EEM$ estimates, we get $\EGW/\EEM < \EGWEMobservedSGRother$ for magnetar burst 2471 and $\EGW/\EEM < \EGWEMobserved$ for magnetar burst 2475.
These $\EGW/\EEM$ ratios are very optimistic, but with aLIGO's increased sensitivity (as will be discussed in \sectionref{sec:discussion}) we can probe more plausible energy ratios for more energetic bursts and giant flares.

It should be noted that we are searching for signals hundreds of seconds long and comparing to energies from short bursts occurring in less than one second.
In most magnetar bursts, the majority of the electromagnetic energy is emitted in the initial burst (occurring in the first second or less) and the energy is smaller in the tail (which when present are order of tens to hundreds of seconds) \cite{article:TheoreticalMagnetarReviewTurolla}.
In some magnetar bursts, the energy in the tail is comparable to or can exceed the initial burst (the energy in the tails of the 1979 and 1998 giant flares were comparable to the initial bursts) \cite{article:TheoreticalMagnetarReviewTurolla}.

Another astrophysical parameter $\EGW$ can be compared to is the estimated magnetic field energy ($E_{\magneticfield}$) of magnetars.
The energy in the magnetic field of magnetars is estimated to be of order $10^{47} \ergs$ if the magnetic field is on order of $10^{15} \gauss$.
This estimate comes from treating the interior magnetic field as a constant $10^{15} \gauss$ field and integrating the calculated magnetic field energy density by the volume of the star, approximated with radius $R \approx 10 \km$. 
The energy fraction is $\EGW/E_{\magneticfield} < \EGWEB$ for the best search sensitivity of this study.
This magnetic energy value could be thought of as a lower limit as it is thought that the internal magnetic field may be stronger than the external field \cite{CorsiOwenEnergyEstimates}.
For an internal field of order $10^{16} \gauss$ (thought to be a plausible value \cite{CorsiOwenEnergyEstimates}), $E_{\magneticfield}$ would be on order of $10^{49} \ergs$, providing a much larger energy reservoir to power GW emission.
\tableref{table:upperLimitsAndResultsSummary} summarizes $\EGW$ for the best search sensitivity for each magnetar burst and their comparisons to these astrophysical parameters.

\begin{table}[h]

\centering

\begin{tabular}{| c | c | c | c |}
	\hline
	\T \B Magnetar Burst & 2469 & 2471 & 2475 \\
	\hline
	\T \B Magnetar & SGR 1806-20 & SGR 1806-20 & 1E 1841-045 \\
	\hline
	\T \B $\EGW$ ($\erg$) & $<\bestULSGR$ & $<\otherULSGR$ & $<\bestUL$ \\
	\hline
	\T \B $\EEM$ $(\erg)$ & Unknown & $7.14\times10^{39}$ & $7.2^{+0.4}_{-0.6} \times 10^{36}$ \\
	\hline
	\T \B $\EGW/\EEM$ & Unknown & $<\EGWEMobservedSGRother$ & $<\EGWEMobserved$ \\
	\hline
	\T \B $\EGW/E_{\magneticfield_{15}}$ & $<\EGWEBSGR{}$ & $<\EGWEBSGRother{}$ & $<\EGWEB{}$ \\
	\hline
\end{tabular}

\caption{Summary of estimated $\EGW$ for the best search sensitivity for three magnetar bursts in S6 and comparisons to possible energy sources.
$E_{\magneticfield_{15}}$ is the energy of a $10^{15} \gauss$ magnetic field and is approximated as $10^{47} \ergs$.}
\label{table:upperLimitsAndResultsSummary}
\end{table}

\section{Future Prospects}\label{sec:discussion}

\begin{figure}[h!]
{\includegraphics[width=\linewidth]{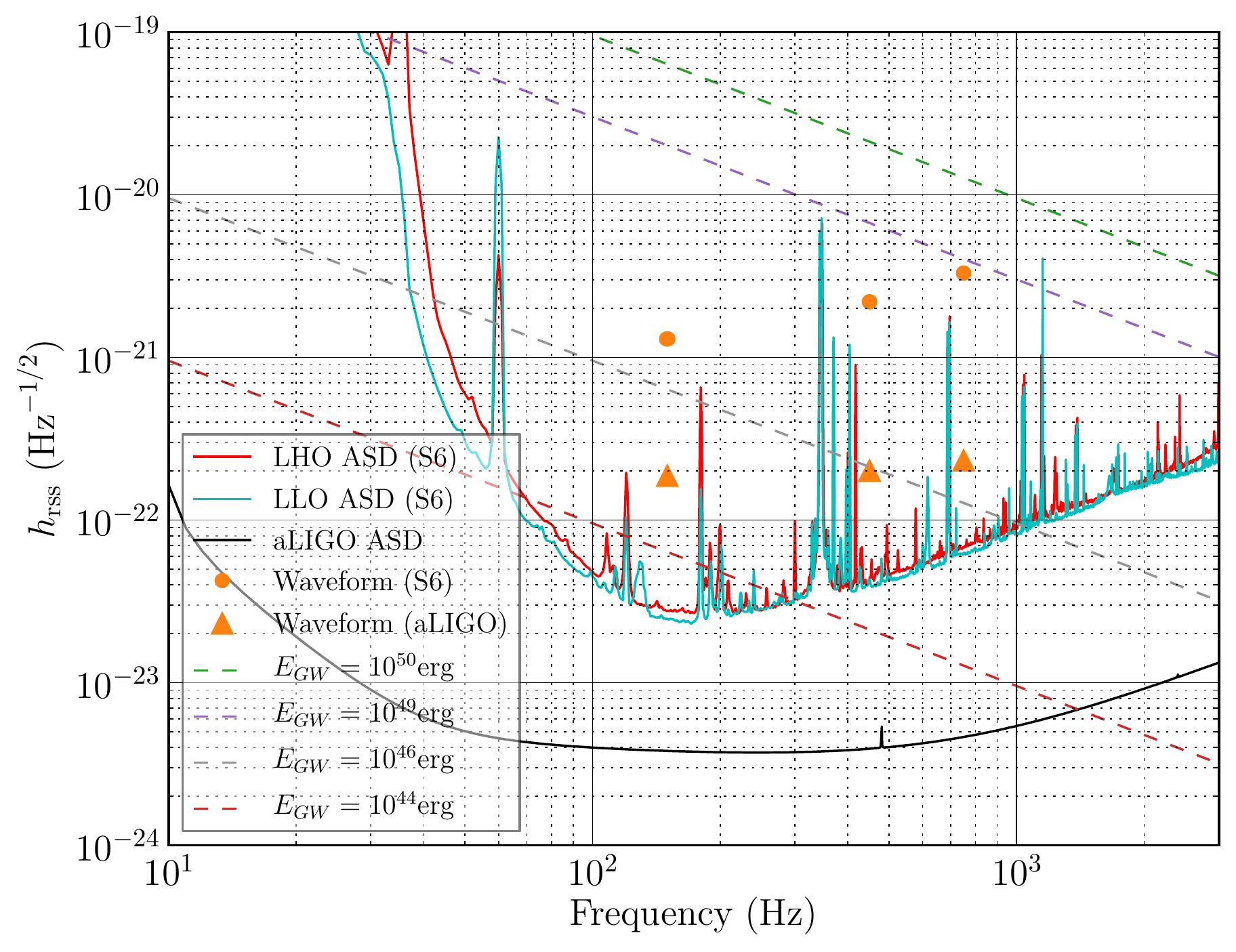}}
\caption{Search sensitivity to half sine-Gaussian signal at frequencies \SI{150}{\Hz}, \SI{450}{\Hz}, and \SI{750}{\Hz} estimated at the time of magnetar burst 2475 (dots). The triangles are an extrapolation of these results to aLIGO design sensitivity. The diagonal lines show the search sensitivity for a source at \SI{8.5}{kpc} from Earth as a function of frequency for three specific energies.
Using aLIGO data, this search method can probe below the energy budgets of possible burst emission mechanisms.
See text for more information.}
\label{fig:hrssestimate}
\end{figure}

This search method can be applied on data for magnetar bursts and giant flares that occur while aLIGO \cite{aLIGOstandardReference} is running.
At design sensitivity \cite{aLIGOdesignCurve}, aLIGO is expected to be $\unsim10$ times as sensitive as initial LIGO, allowing it to detect GWs with energies $\unsim 100$ times smaller.
The search sensitivity of this method estimated with LIGO data from S6 can be extrapolated to aLIGO design sensitivity, as shown in \figureref{fig:hrssestimate}.
This extrapolation suggests 
that the strong magnetic fields of magnetars could power detectable GWs at the distances of known magnetars (1.6--\SI{62.4}{kpc} \cite{mcgillMagnetarList}) if the GW energies are of similar energies to the electromagnetic emissions.

In \figureref{fig:hrssestimate}, the $\hrss$ of the estimated search sensitivity for waveforms with $\tau = \SI{400}{\s}$ are plotted, as are the $\hrss$ for the extrapolated aLIGO search sensitivities to these waveforms.
We also plot the amplitude spectral density (ASD) for LHO and LLO from the on-source window of magnetar burst 2471.
The aLIGO design sensitivity \cite{aLIGOdesignCurve} ASD is also plotted.
Diagonal lines are plotted for the $\hrss$ (as calculated in \appendixref{upper_limits}) for specific waveform energies from a source at \SI{8.5}{\kpc} as a function of frequency; the waveform energies are $10^{50} \ergs$ (the highest energy budget for crust cracking \cite{CorsiOwenEnergyEstimates}),  $10^{49} \ergs$ (the energy budget for hydrodynamic deformation \cite{CorsiOwenEnergyEstimates}), $10^{46} \ergs$ (the same order of magnitude as the 2004 giant flare), and $10^{44} \ergs$ (the order of magnitude of the other giant flares).

We extrapolate aLIGO's search sensitivity using the ratio of the LHO and LLO ASDs in SGR trigger 2475's on-source window to aLIGO's design sensitivity at the central frequency of each waveform.
This extrapolation obtains waveform energies in the range of the observed electromagnetic energies of magnetar bursts.
The narrow-band nature of the injected waveforms means extrapolating to aLIGO is relatively simple, since, to first order, the differences to the signal arise from the noise amplitude rather than the spectral shape of the detector response.
At aLIGO design sensitivity, the estimated minimum detectable waveform energy for this study of $\EGW =  \bestUL \erg$ could have been $\EGW = \bestULaLIGO \erg$.
Additionally, there are multiple magnetars with distances of order \SI{2}{kpc}, and one as close as 1.6 kpc \cite{mcgillMagnetarList}.
If a burst from a magnetar 1.6 kpc away occurs during aLIGO, the estimated minimum detectable waveform energy could be $\EGW = \unsim\bestULaLIGOclose \erg$, the same order of magnitude as some stronger intermediate bursts.

The estimated search sensitivity for aLIGO can be compared with the electromagnetic energy radiated during giant flares.
If a giant flare with $\EEM \approx 10^{44} \ergs$ occurred, the ratio of GW energy to emitted electromagnetic energy for the estimated search sensitivity of a magnetar at \SI{1.6}{kpc} would be
$\EGW/\EEM < \bestULaLIGOcVsBurst$.
For an event such as the 2004 giant flare with $\EEM = \unsim 1.7 \times 10^{46} (d/\SI{8.7}{\kpc})^2 \ergs$, where $d$ is the distance to SGR 1806-20, this ratio would be $\EGW/\EEM < \bestULaLIGOcVsGF$.
These comparisons to estimated aLIGO search sensitivity are summarized in \tableref{table:upperLimitsaLIGO}.
With these energy ratios, a GW with high $\EGW/\EEM$ would either be detected or ruled out.

\begin{table}[h]

\centering

\begin{tabular}{| c | c |}
	\hline
	\T \B Source Magnetar & 1E 1841-045 \\
	\hline
	\T \B $\EGW$ & $\bestUL \erg$ \\
	\hline
	\T \B Estimated $\EGW$ for aLIGO & $\bestULaLIGO{} \erg$ \\
	\hline
	\T \B $\EGW{}_{, \mathrm{aLIGO}}$ for 1.6 kpc & $\bestULaLIGOclose{} \erg$ \\
	\hline
	\T \B $\EGW{}_{, \mathrm{aLIGO}, 1.6 \mathrm{kpc}}/\EEM$ ($10^{44} \ergs$) & $\bestULaLIGOcVsBurst{}$ \\
	\hline
	\T \B $\EGW{}_{, \mathrm{aLIGO}, 1.6 \mathrm{kpc}}/\EEM$ ($2 \times 10^{46} \ergs$) & $\bestULaLIGOcVsGF{}$ \\
	\hline
\end{tabular}

\caption{Summary of estimated $\EGW$ for the best search sensitivity and equivalent $\EGW$ for aLIGO. Includes comparisons to electromagnetic energy levels of giant flares.}
\label{table:upperLimitsaLIGO}
\end{table}

The sensitivity estimates of this search method can also be compared to energy budgets of possible burst and giant flare emission mechanisms.
As discussed in \sectionref {sec:magnetarburstsandgws}, hydrodynamic deformation could generate up to $10^{49}\ergs$ \cite{article:Ioka, CorsiOwenEnergyEstimates}, and crust cracking could generate up to $10^{46} \ergs$ in the case of a normal neutron star and \mbox{$10^{49}$--$10^{50} \ergs$} if the star contains solid quark matter \cite{CorsiOwenEnergyEstimates}.
If these mechanisms are responsible for magnetar bursts or giant flares, they could provide sufficient energy for detectable GWs.

Detection of GWs related to magnetar bursts is an interesting prospect.
These sensitivity estimates suggest running this search method on data from aLIGO could probe significantly below the energy budgets of crust cracking and hydrodynamic deformation (see \figureref{fig:hrssestimate}), providing sensitivity to GWs with energies comparable to the electromagnetic energies of intermediate bursts and giant flares.
More detectors are expected to join the global network of ground based GW detectors, including Virgo \cite{aVirgo}, KAGRA \cite{kagraPaper}, and LIGO-India \cite{LIGOIndia}, and hence can improve the search sensitivity.
It is straightforward to extend this search method for a network of more than two detectors \cite{STAMPmethodsPaper}.
The results of a search using this method could place constraints on the energy released through mechanisms such as crust cracking and hydrodynamic deformation, and constrain the possible equations of state of magnetars, thereby increasing our knowledge of the astrophysics of magnetars.

\acknowledgments
We thank the LIGO Scientific Collaboration for the use of LIGO data in this study.
This research has made use of data, software and/or web tools obtained from the LIGO Open Science Center (https://losc.ligo.org), a service of LIGO Laboratory and the LIGO Scientific Collaboration. LIGO is funded by the U.S. National Science Foundation.
This work was supported by NSF PHY-1607336.
MC is supported by National Science Foundation Graduate Research Fellowship Program, under NSF grant number DGE 1144152.
ET is supported through ARC FT150100281 and CE170100004.

\appendix

\section{Estimating $\hrss$ and $\EGW$}\label{upper_limits}

The root sum square strain of a signal $h$ is defined as
\begin{equation}\label{hrss}
h_{rss} = \sqrt{\definiteIntegral{-\infty}{\infty}{|h|^2}{t}}\;,
\end{equation}
where $|h|^2 = |h_+|^2 + |h_\times|^2$. For the second half of a sine-Gaussian waveform of the form in \eqref{fullHplusSG} and \eqref{fullHcrossSG}, $\hrss$ is:
\begin{multline}\label{hrssQ}
h_{\mathrm{rss}} = \frac{h_0}{4\pi^{1/4}}
\sqrt{\frac{Q}{f_0}}
\left[
\left(
\frac{\left(1 + \cos^2\iota\right)^2}{4} + \cos^2{\iota}
\right) \right. \\
\left. + \left(\frac{\left(1 + \cos^2\iota\right)^2}{4} - \cos^2{\iota}\right)
\euler^{-Q^2}
\right]^{1/2}\;.
\end{multline}
Here $Q$ is the quality factor $Q = \sqrt{2} \pi \tau f_0$. For high $Q$, this is given by
\begin{equation}\label{hrssQhighQ}
h_{\mathrm{rss}}^{\mathrm{HQ}} \approx \frac{h_0}{4\pi^{1/4}}
\sqrt{\frac{Q}{f_0}}
\sqrt{
\left(
\frac{\left(1 + \cos^2\iota\right)^2}{4} + \cos^2{\iota}
\right)
}\;.
\end{equation}

The GW energy for a half sine-Gaussian is given by
\begin{equation}\label{EGWExact}
E_{\mathrm{GW}} =
h_0{}^2 r^2 Q 
f_0
\frac{c^3\pi^{3/2} }{20 \, G} 
\left[
1
+ \frac{1}{2Q^2} \left(1
+ \frac{1}{6}\euler^{-Q^2}
\right)
\right]\;.
\end{equation}
For high $Q$, \eqref{EGWExact} can be approximated as:
\begin{equation}\label{EGWHQ}
E_{\mathrm{GW}}^{\mathrm{HQ}} \approx
\frac{c^3\pi^{3/2} }{20 \, G}
h_0{}^2 r^2 Q f_0\;.
\end{equation}
For the $h_{\mathrm{rss}}$ and $E_{\mathrm{GW}}$ of a full sine-Gaussian waveform, multiply \eqref{hrssQ} and \eqref{hrssQhighQ} by $\sqrt2$ and multiply \eqref{EGWExact} and \eqref{EGWHQ} by 2. \eqref{hrssQ}, \eqref{hrssQhighQ}, \eqref{EGWExact} and \eqref{EGWHQ} are calculated in detail in Appendix B of \cite{RQJthesis}.

\section{Calibration Error}\label{cal_err}

The measured cross-correlated signal $h_{0 \mathrm{m} \, 1} h_{0 \mathrm{m} \, 2}$ has uncertainties due to calibration error.
The detector differential arm lengths are measured in voltage counts.
Measurements must be made on different systems within the detector to measure the conversion from these voltage counts to actual strain values. The conversion function is known as the calibration. Errors in the calibration propagate into error in the measured value of the strain.
The calibration errors can be grouped into an overall scaling error $A$, a frequency dependent amplitude error, a frequency dependent phase error $\delta$ and a timing error \cite{S6Calibration}. 
The timing error is effectively a linear frequency dependent phase error.

The effect of the calibration error on the measured strain value can be estimated with two extremes. The first extreme would be the most sensitive to a potential signal (the scaling error is $\delta A_-$, the phase errors of each detector cancel out and the timing error is zero):
\begin{multline}\label{calibrationErrorBest}
\frac{h_{0\, 1} h_{0\, 2}}{h_{0 \mathrm{m} \, 1} h_{0 \mathrm{m} \, 2}} = A_1 A_2 \\ \times \left(1-\sqrt{\delta A_{-1}{}^2 + \delta A_{-2}{}^2 +  \left|\Delta h_1(t)\right|^2 +  \left|\Delta h_2(t)\right|^2} \right)\;.
\end{multline}
In the other extreme, the calibration error leads to an underestimation of the strength of a potential signal (the scaling error is $\delta A_+$, the phase error of the detectors would be of the same sign with maximum timing error):
\begin{multline}\label{calibrationErrorWorst}
\frac{h_{0\, 1} h_{0\, 2}}{h_{0 \mathrm{m} \, 1} h_{0 \mathrm{m} \, 2}} = A_1 A_2 \\ \times \frac{1+\sqrt{\delta A_{+1}{}^2 + \delta A_{+2}{}^2 +  \left|\Delta h_1(t)\right|^2 +  \left|\Delta h_2(t)\right|^2}}{\cos{\left(2\pi f 45 \mu \mathrm{s} + \delta_1(t) + \delta_2(t) \right)} }\;.
\end{multline}

For detailed derivations of \eqref{calibrationErrorBest} and \eqref{calibrationErrorWorst} and specific numbers on the different calibration errors, see Appendix C of Ref.~\cite{RQJthesis}.
The calibration of S6 is discussed in detail in Ref.~\cite{S6Calibration}.
In order to find a conservative search sensitivity, the calibration error in \eqref{calibrationErrorWorst} was assumed.
To account for the calibration error, $\EGW$ was multiplied by the ratio in \eqref{calibrationErrorWorst}. Similarly, $\hrss$ was multiplied by the square root of this ratio.

\bibliography{bibliography}

\end{document}